\documentclass[aps,floatfix,twocolumn,superscriptaddress]{revtex4-2}

\usepackage{amsmath} 
\usepackage{amssymb}
\usepackage[normalem]{ulem}
\usepackage{mathtools} 
\usepackage[T1]{fontenc}
\usepackage{fancyhdr} 
\usepackage{newtxtext,newtxmath}
\usepackage[usenames, dvipsnames]{color}
\usepackage{verbatim}
\usepackage{graphicx}                      
\usepackage{epstopdf}               
\graphicspath{{./figures/}}                
\usepackage[colorlinks=true, linkcolor=Blue, allcolors=Blue]{hyperref}
\usepackage{blindtext}
\usepackage{natbib}

\usepackage{hyperref}
\usepackage{nameref}
\usepackage[usenames, dvipsnames]{color}

\def\MODIF#1{{\textcolor{black}{#1}}}

\begin{document}

\title{\MODIF{Backscattering reduction in a sharply bent water wave channel}}
\author{S. Kucher} 
\affiliation{Laboratoire de Physique et M\'ecanique des Milieux H\'et\'erog\.enes, UMR CNRS 7636, ESPCI-Paris, Universit\'e PSL, Sorbonne Universit\'e, Universit\'e Paris Cit\'e, 75005 Paris, France} 

\author{A. Ko{\'{z}}luk}
\affiliation{Laboratoire de Physique et M\'ecanique des Milieux H\'et\'erog\.enes, UMR CNRS 7636, ESPCI-Paris, Universit\'e PSL, Sorbonne Universit\'e, Universit\'e Paris Cit\'e, 75005 Paris, France} 
\affiliation{Institute of Aeronautics and Applied Mechanics, Warsaw University of Technology, 00-665 Warsaw, Poland}

\author{P. Petitjeans} 
\affiliation{Laboratoire de Physique et M\'ecanique des Milieux H\'et\'erog\.enes, UMR CNRS 7636, ESPCI-Paris, Universit\'e PSL, Sorbonne Universit\'e, Universit\'e Paris Cit\'e, 75005 Paris, France} 

\author{A. Maurel} 
\affiliation{Institut Langevin, UMR CNRS 7587, ESPCI-Paris, 75005 Paris, France}

\author{V. Pagneux} \email[Corresponding author: ]{vincent.pagneux@univ-lemans.fr}
\affiliation{Laboratoire d'Acoustique de l'Universit\'e du Maine, UMR CNRS 6613, 72085 Le Mans, France}

\begin{abstract} 

We study theoretically and experimentally how to reduce the backscattering of water waves in a channel with multiple turns.  We show that it is not only possible to cancel backscattering but also to achieve a remarkable transmission in such geometries. In order to avoid the reflection that naturally arises at each turn of the waveguide, an anisotropic metamaterial made of closely-spaced thin vertical plates is used. The efficiency of the metamaterial arrangement depends only slightly on the frequency of the incident wave, as long as its wavelength is much larger than the periodicity of the array.
This phenomenon is not restricted only to water wave channels but also applies to any type of waves with Neumann boundary conditions. 
\newline \newline
DOI: (to be specified) 
\end{abstract}

\maketitle 
\thispagestyle{fancy}

\section{Introduction}

The use of metamaterials for wave manipulation has generated considerable interest in recent years, both in electromagnetics as well as in acoustics and surface waves \cite{bliokh2008colloquium, cui2010metamaterials, Craster2013Acoustic, romero2019fundamentals}. Metamaterials typically consist of a periodic structure whose characteristic length is much smaller than the considered wavelength and allow to modify the natural propagation of waves. They are able to generate, for example, invisibility cloaking \cite{Schurig2006, farhat2008broadband, Zou2019}, wave absorption \cite{Landy2008, romero2016perfect, monsalve2019, de2021attenuating} or wave shifting \cite{Chen2008, Berraquero2013, Xie2014, Wei2015}, among other effects.

In particular, in the context of water waves, one type of metamaterial commonly used to control wave propagation is an array of rigid plates forming a subwavelength grating. This type of systems have been studied both experimentally \cite{Berraquero2013, Maurel2017a} and theoretically \cite{Porter2021, porter2022water, wilks2022rainbow, huang2023water}. They have been found to possess Brewster angle-type behavior, in which backscattering is considerably reduced and near-zero reflection values are achieved for a wide range of frequencies \cite{Alu2011, Akozbek2012, Norris2015, Pham2020}, without relying on resonant phenomena. \MODIF{When it is not zero reflection  for a wide frequency band that is sought but that the objective is rather the stability of transparency with angle, note that wide angle transparent anisotropic metamaterial have been proposed  for elctromagnetic waves \cite{he2018magnetoelectric, chu2022extremely}.}

In order to facilitate the modelling of the properties of gratings with a subwavelength structure, homogenization techniques have been developed \cite{Maurel2013,  Norris2015, Maurel2017a,Marigo2017, Pham2020}. These techniques replace the perforated domain with a homogeneous and anisotropic one. At low enough frequencies, under the homogenization regime approximation, they are capable of accurately predict the behavior of the system, including both the decrease in backscattering at Brewster angle and in resonances. If the plates constituting the grating occupy the entire depth of the fluid, going from the bottom to the surface, since the space between them is smaller than the incident wavelength, the energy flow is forced in one direction only. Of particular interest is the recent work of Porter \cite{Porter2021}, in which he showed that if the thickness of the plates is infinitely thin, then the reflection is exactly zero (i.e., without making use of the homogenization regime approximation).

Naturally, there has been an interest in performing experiments on shifting waves using a subwavelength grating. For example, reflectionless waveguides with an angle up to $\pi/6$ have been achieved for water waves \cite{Berraquero2013}, using theory of transformation media to design a metabathymetry with anisotropic properties. In the context of acoustics, a beam shifter capable of exhibiting a high transmission at resonant frequencies and at a Brewster-like incidence angle has been realized \cite{Wei2015}. 

In the present study, we show theoretically and experimentally a broadband backscattering reduction in a \MODIF{sharply bent} water wave channel. It is achieved owing to an array of vertical surface piercing thin plates with a perpendicular angle with respect to the incident wave, as shown in Fig. \ref{fig:setup}. The effect is broadband because the spacing between the plates is much smaller than the incident wavelength, resulting in an effective medium without impedance mismatch between the inside and the outside. 
The paper is organized as follows. In Sec. \ref{sec:channel} we characterize the reflection generated by the \MODIF{bending} of a waveguide. The use of a plate-array metamaterial to reduce the backscattering after each turn is proposed in Sec. \ref{sec:metamaterial}.
We study the reflection coefficient given by a plane wave incidence on a plate-array grating both in an infinite domain and in a \MODIF{bent} waveguide. We also propose a homogenized model to replace the subwavelength grating by an effective medium. Finally, in Sec. \ref{sec:experiments} we show an experimental realization of this system, exploring its capabilities in terms of frequency.

\section{\label{sec:channel}\protect\MODIF{Sharply bent water wave channel}}

\begin{figure}[!t] 
    \centering
    \includegraphics[width=\linewidth]{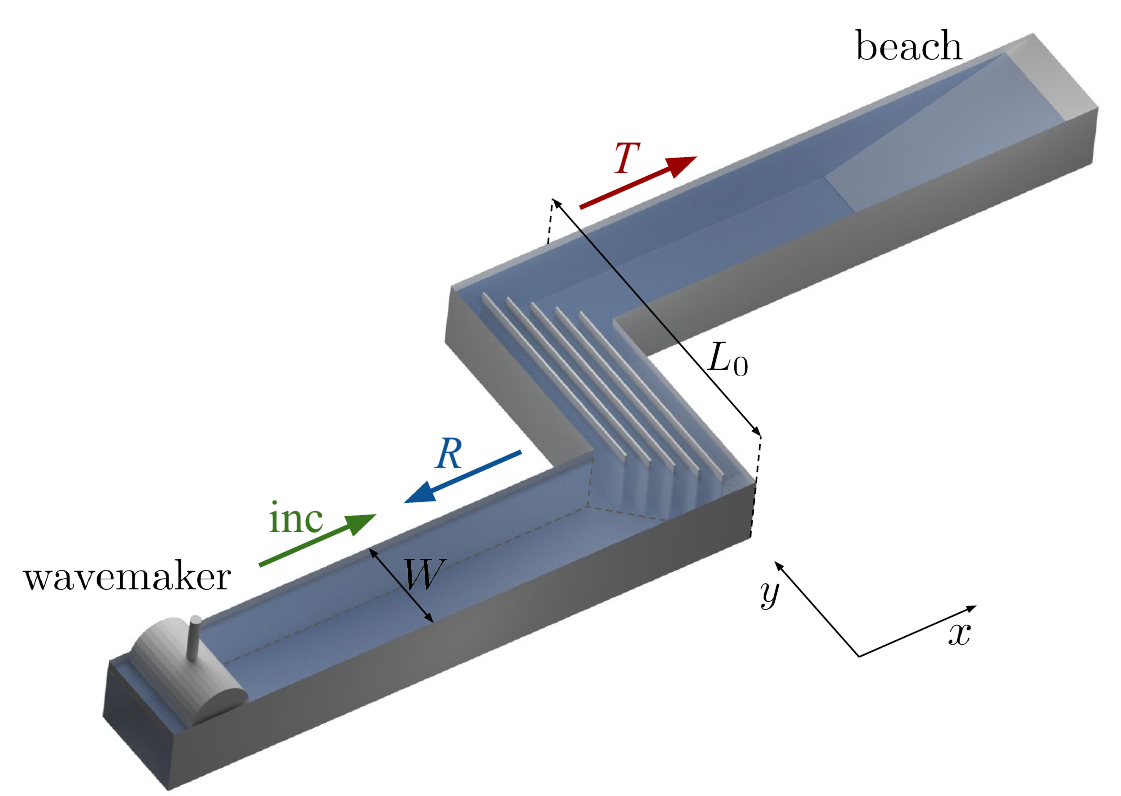}
    \caption{
    Scheme of the experimental setup: a channel of width $W$ with a dislocation of length $L_0$, where the plate-array metamaterial is placed. An incident wave is reflected and transmitted with scattering coefficients $R$ and $T$.
    } 
    \label{fig:setup} 
\end{figure}

The system under consideration is depicted in Fig. \ref{fig:setup}. Surface waves are generated at the beginning of the channel, which undergoes a perpendicular \MODIF{turn} of length $L_0$ and then continues in a direction parallel to the first section. At the end of the channel an absorbing beach avoids reflections. We first consider this geometry without metamaterial to characterize the amount of reflection it generates. Thereafter we will show, in the next section, that it is possible to reduce the backscattering generated by the \MODIF{turn} by adding a plate-array metamaterial in the central region that allows to achieve high values of transmission as well.

The free surface elevation $\eta(x,y)$ with time harmonic dependence $e^{-i\omega t}$ is governed by the 2D Helmholtz equation
\begin{equation}
	(\Delta + k^2)\eta(x,y)  = 0, 
\label{eq:helmholtz}	
\end{equation}

\noindent where $k$ is the wavenumber, along with Neumann boundary conditions on the vertical walls $\partial_n\eta = 0$. 
\MODIF{This equation is valid for irrotational and incompressible flows in geometries with a flat bottom and vertical walls in the harmonic regime, under the linear approximation.}


We consider waves that are generated for frequencies below the cut-off frequency of the waveguide of width $W$, i.e., $kW~<~\pi$, which only allows the propagation of the planar mode. Under the wide spacing approximation, it is then possible to construct a 1D model that takes into account the scattering coefficients of each perpendicular turn individually, along with 1D propagation following the waveguide (see Appendix A). It can be noticed that this system is analogous to a Fabry-Perot interferometer, in which waves pass through a cavity made up of two parallel reflecting surfaces separated by a distance $L = L_0 - W$. Following this 1D model, the absolute value of the reflection coefficient of the \MODIF{bent} channel is given by
\begin{equation}
	|R|^2 = \frac{F\sin^2(\frac{\delta}{2})}{1+F\sin^2(\frac{\delta}{2})},
\label{eq:fabryperot}	
\end{equation}

\noindent defining $F = 4r_0^2/(1-r_0^2)^2$ and $\delta = 2kL+2\varphi$, where $r_0$ and $\varphi$ are respectively the modulus and the phase of the reflection coefficient ($r~=~r_0 e^{i \varphi}$) for one single perpendicular turn (see Appendix B).

\begin{figure}[!t] 
    \centering 
    \includegraphics[width=\linewidth]{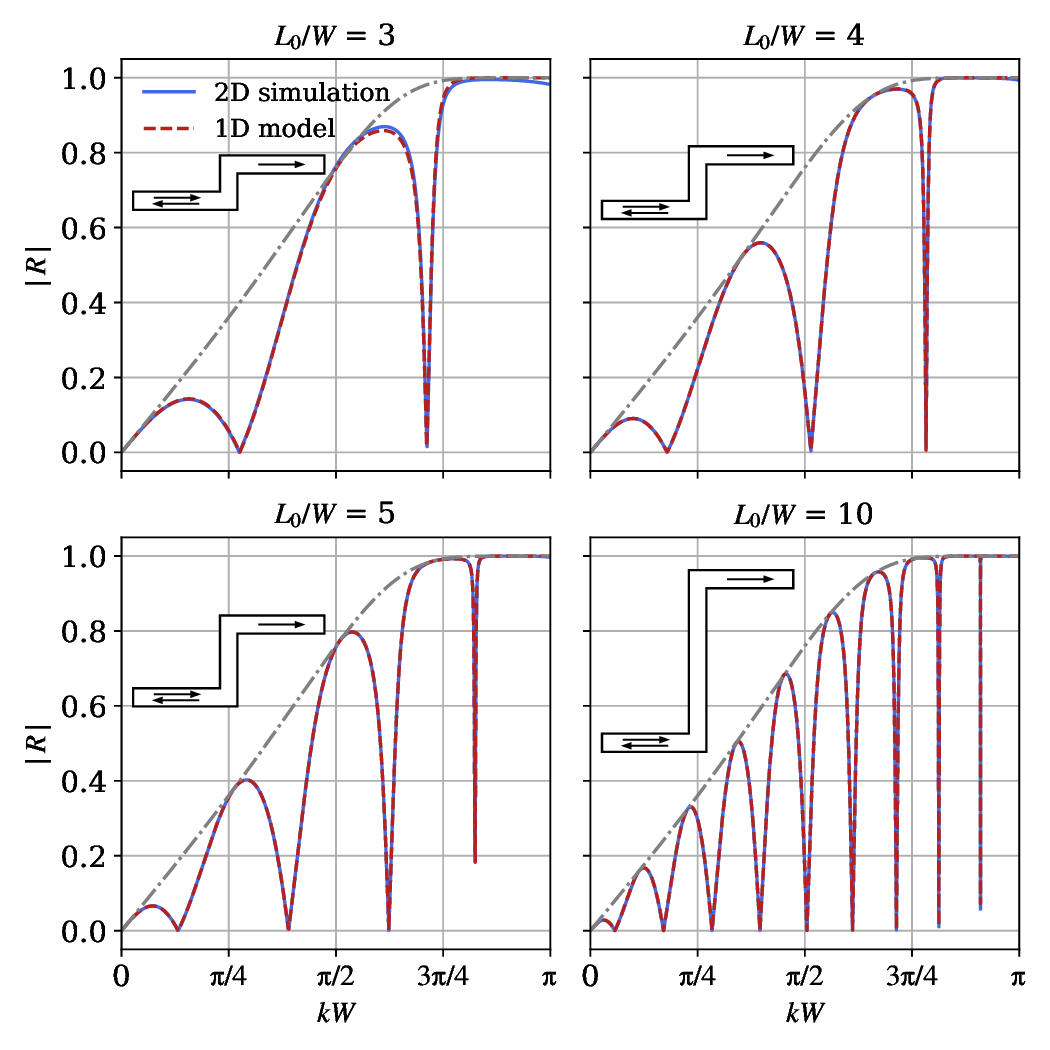}
    \caption{
	    Reflection coefficient for different ratios of $L_0/W$ without the metamaterial. The solid lines (blue) exhibit the result of a 2D simulation of the full channel, the dashed (red) lines indicate the Fabry-Perot resonances and the dotted (gray) lines shows the envelope of $|R|$, obtained from eq. \ref{eq:fabryperot}.
    } 
    \label{fig:fabry-perot} 
\end{figure}

Clearly, backscattering suppression is achieved when $\sin^2 (\delta/2) = 0$, i.e., when $kL + \varphi = m\pi$ ($m \in \mathbb{Z}$). Additionally, the envelope of this expression can be obtained by imposing $\sin^2 (\delta/2) = 1$ and is given by $|R|^2 = F/(1+F) $, which depends only on $r_0$. We may note that the absolute value of the reflection coefficient for one corner yields the shape of the envelope, while its phase provides the position of the resonances, as we will see in Fig. \ref{fig:fabry-perot}.

In order to study the dependence of the reflection coefficient $R$ on the distance $L_0$ of the perpendicular \MODIF{turn}, Fig. \ref{fig:fabry-perot} shows the values of $|R|$, for different ratios of $L_0/W$, obtained from  the 1D model (eq. \ref{eq:fabryperot}) as well as the ones computed numerically by solving the Eq. \ref{eq:helmholtz} in the full 2D geometry by finite element method (MATLAB PDEtool). There are two behaviors that are worth noting when the ratio $L_0/W$ is increased. Firstly, the number of resonances becomes consequently higher, as expected from eq. \ref{eq:fabryperot}. Secondly, the 1D model agrees very well with the full 2D simulation and, unsurprisingly under the wide spacing approximation, this agreement is better and better when the two perpendicular corners move away from each other since the near field effects decrease. 

Regardless of the $L_0/W$ ratio, the general tendency of the reflection coefficient is the same. It is zero at resonance frequencies, while its upper envelope $\sqrt{F/(1+F)}$  does not depend on $L_0/W$. In the following, we will add a plate-array metamaterial in order to reduce the backscattering of the \MODIF{turn} and, without loss of generality, we will work at $L_0/W$ = 3.2. 

\section{\label{sec:metamaterial}Plate-array metamaterial}

With the aim of reducing the reflection that naturally arises at the \MODIF{turn}, we are going to use an anisotropic metamaterial that significantly diminishes the reflection at each turn. The metamaterial consists of an array of thin parallel and closely spaced plates of length $L = L_0-W$ vertically placed in the central region, as shown in Fig. \ref{fig:setup}. The vertical plates are surface piercing, spanning from the channel bottom to above the fluid surface. 

\subsection{Reflection on an infinite grating}

\begin{figure}[!t] 
    \centering 
    \includegraphics[width=\linewidth]{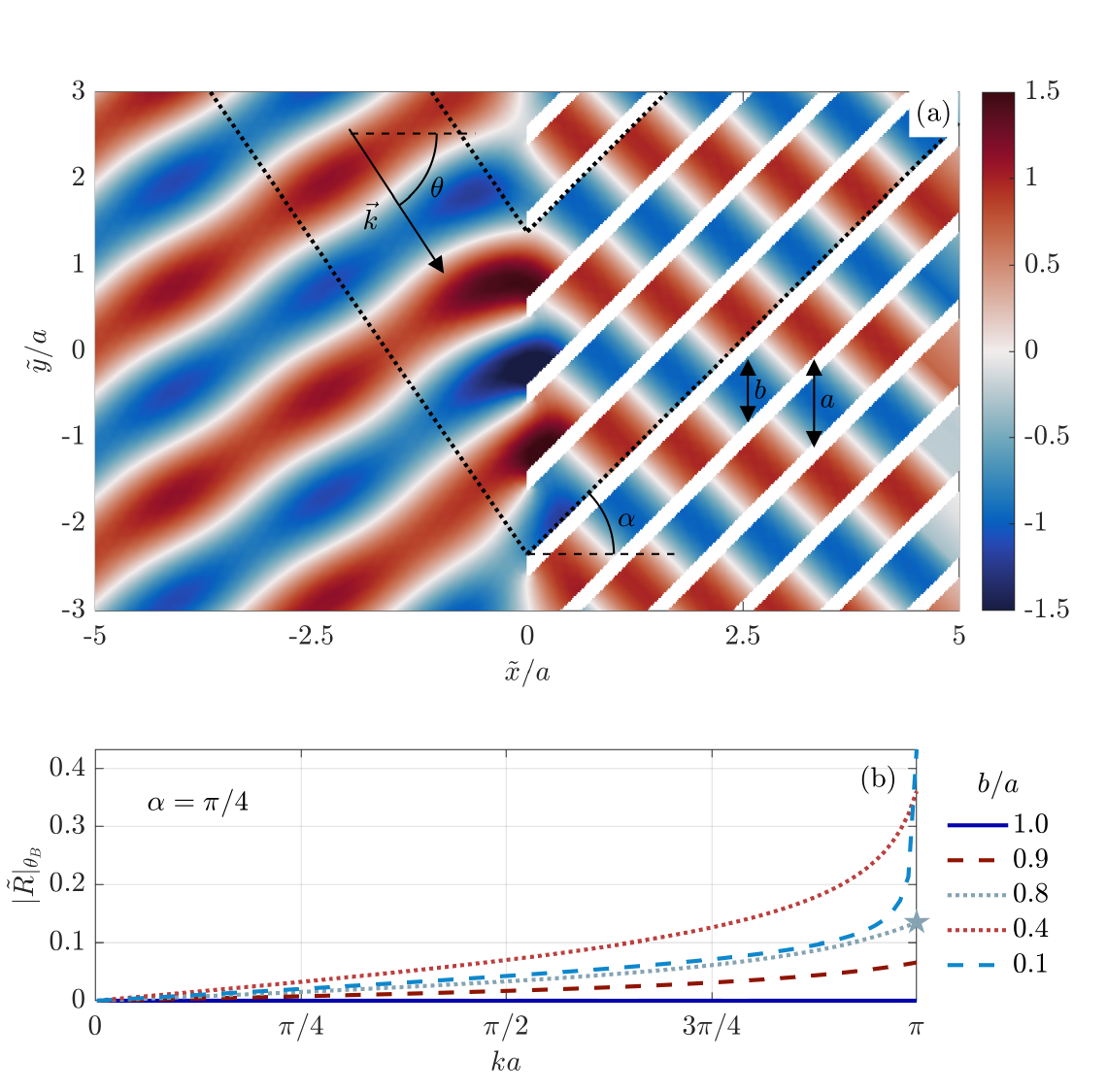}
    \caption{(a) Real part of the surface elevation field $\eta (x,y)$ for a plane wave incidence with wavenumber $k$ and incident angle $\theta$ over a semi-infinite array of tilted plates with a periodicity $a$ and inclination angle $\alpha = \pi/4$. $b$ is the spacing between the plates. Here $b/a = 0.8$, $ka = \pi$ and $\theta = \theta_B$. The dotted lines indicate where the channel walls would be if we confined this system as in Fig. \ref{fig:setup}. (b) Reflection coefficient as a function of $ka$ for this geometry with $\alpha = \pi/4$, for different values of the filling ratio $b/a$, for a plane wave incidence at the Brewster angle $\theta_B$ (eq. \ref{eq:brewster}). The star indicates the point that corresponds to the surface showed in (a). 
    } 
    \label{fig:comsol_surface} 
\end{figure}

First, to assess the reflection reduction capabilities, it is necessary to determine the characteristics of the plate-array metamaterial, namely, the thickness of the plates and the spacing between them. In order to understand the effect of the plate thickness on the scattered fields, we begin by considering the problem of a plane wave incidence on an infinite periodic grating made of parallel inclined plates, as shown in Fig. \ref{fig:comsol_surface}. The periodic array is composed of plates with periodicity $a$ separated by a distance $b$, making an angle $\alpha$ with respect to the $\tilde{x}$-axis \cite{note_axis_grating}. We consider a plane wave incidence with incident angle $\theta$: for $ka \le \pi$, no higher order modes can propagate, and there is only reflection with coefficient $\tilde{R}(k, \theta)$ at the specular angle by the grating.
 
If the wavelength is much smaller than the periodicity of the array, such that $ka \ll 1$, the hypothesis of the homogenization regime is satisfied. In this case, it is possible to obtain a Brewster like behavior, i.e., a zero of reflection for an incidence angle $\theta_B$ given by 
\begin{equation}
	\theta_B = \cos^{-1}\Big(\frac{b}{a}\cos\alpha\Big), 
	\label{eq:brewster}
\end{equation}
\noindent in agreement with previous works \cite{Norris2015}. It is worth mentioning that when the plates have no inclination ($\alpha = 0$), we recover the expression for the Brewster angle as it is usually found for subwavelength gratings \cite{Maurel2013}. 

Since we use the approximation $ka \ll 1$ to obtain the expression for $\theta_B$ (eq. \ref{eq:brewster}), it is expected that the reflection is not exactly zero. We examine then the exact reflection of a plane wave at the Brewster angle $\theta_B$, for different values of the filling ratio of water $b/a$. Results are displayed in Fig. \ref{fig:comsol_surface} for the case of $\alpha=\pi/4$. We utilized COMSOL Multiphysics to conduct these numerical simulations. In Fig. \ref{fig:comsol_surface}(a) an example of the solution for $b/a = 0.8$ is presented; even at the Brewster angle, there is a small amount of reflection, due to the deviation from the homogenization theory. To get a quantitative evaluation of these small reflection values at Brewster angle, Fig. \ref{fig:comsol_surface}(b) illustrates the absolute value of the reflection coefficient $|\tilde{R}|_{\theta_B}$ obtained for various configurations, changing the thickness of the plates. As expected, $|\tilde{R}|_{\theta_B}$ increases with $ka$. But, surprisingly, this trend is not monotonous with $b/a$, with a maximum around $b/a \approx 0.4$. Incidentally, we verified that this tendency is independent of the value of $\alpha$.  

The case where $b=a$ is of particular interest because it corresponds to exact zero reflection up to $ka=\pi$ for the Brewster angle. This property can be explained by two arguments \cite{Norris2015}: i) the invisibility of infinitesimally thin plates when $\vec{k}$ is parallel to the plates, and ii) the reciprocity implying $\tilde{R}(\theta) = \tilde{R}(-\theta)$. 
In the following, we will work with the thinnest plates achievable (since a zero thickness is unattainable in an experimental setup). Moreover, the angle $\alpha=\pi/4$ that we have considered in the results yields a right angle between the incident and transmitted waves for the Brewster case, that corresponds to the turn of $\pi/2$ in the \MODIF{bent} waveguide.

\subsection{Reflection on a grating in the \MODIF{bent} waveguide}

\begin{figure}[!t] 
    \centering 
    \includegraphics[width=\linewidth]{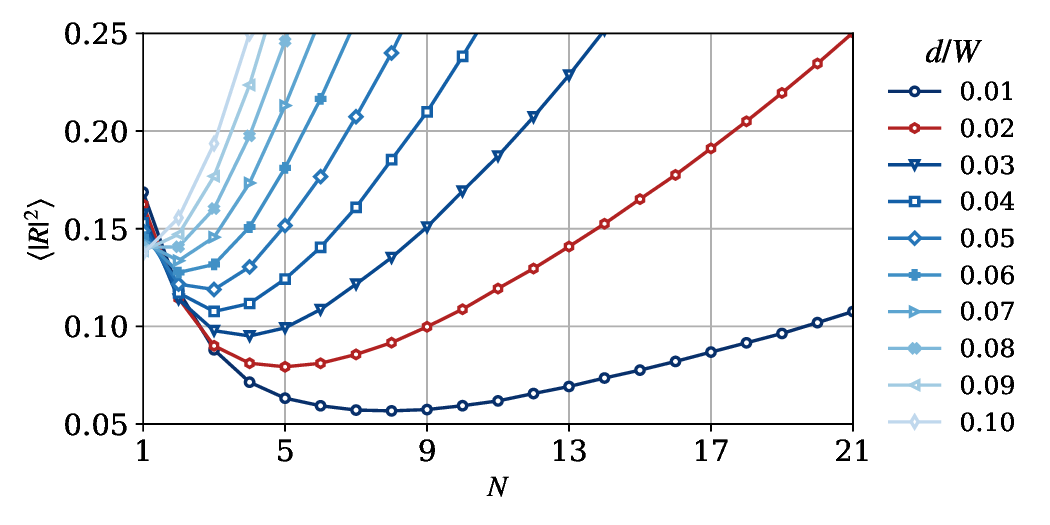}
    \caption{
	    Mean (over all the frequencies below $f_{\text{cut-off}}$) of the absolute value of the reflection coefficient as a function of the number of plates for different values of thickness. The thickness that we chose to perform experiments is indicated in red.
    } 
    \label{fig:optimal_n_multiple_thickness} 
\end{figure}

Bringing our focus back to the primary objective, i.e., to reduce the backscattering in a \MODIF{sharply bent} water wave channel, we consider plates with a finite thickness $d$ and we confine the system into a waveguide of width $W$ with two perpendicular turns, each turn being constructed as indicated with dotted lines in Fig. \ref{fig:comsol_surface}(a). Since we are going to perform experiments, we chose the thickness of the plates to be $d = 0.02W$ so that the reflection is minimized and the plates are rigid enough.
Now the reflection is not exactly zero for two reasons: firstly, the plates have a non-zero thickness and secondly, there are edge effects because the channel walls stop the periodicity. \MODIF{The case of infinitesimally thin plates in a waveguide is considered in Appendix C to illustrate these edge effects.} In the following, we will evaluate the impact of the plate thickness on $|R|$.

We performed numerical simulations taking a number $N$ of plates and computing the mean value of $|R|^2$ over all the frequencies below $f_{\text{cut-off}}$, for different values of thickness $d/W \in [0.01, 0.10]$. The obtained curves are shown in Fig. \ref{fig:optimal_n_multiple_thickness}. We can observe that, as expected, the mean reflection diminishes with the plate thickness. The number of plates also plays a role in the reflection coefficient. For a small number of plates, the hypothesis $ka \ll 1$ is not satisfied, while for a large value of $N$, the high density of plates creates a barrier within the channel. There is therefore an optimal number of plates for each thickness for which the minimum of $|R|$ is achieved. For the chosen value $d/W = 0.02$ (red curve in Fig. \ref{fig:optimal_n_multiple_thickness}), the optimal number of plates is $N = 5$. In this case, despite having non-zero thickness plates (rigid enough), the reflection coefficient is globally (broadband in frequency) small, not so far from the ideal case of zero thickness plates for an infinite grating discussed in the previous section.

\subsection{Homogenized model}

\begin{figure}[!t] 
    \centering 
    \includegraphics[width=\linewidth, trim={0.2cm 0 0.3cm 0.2cm},clip]{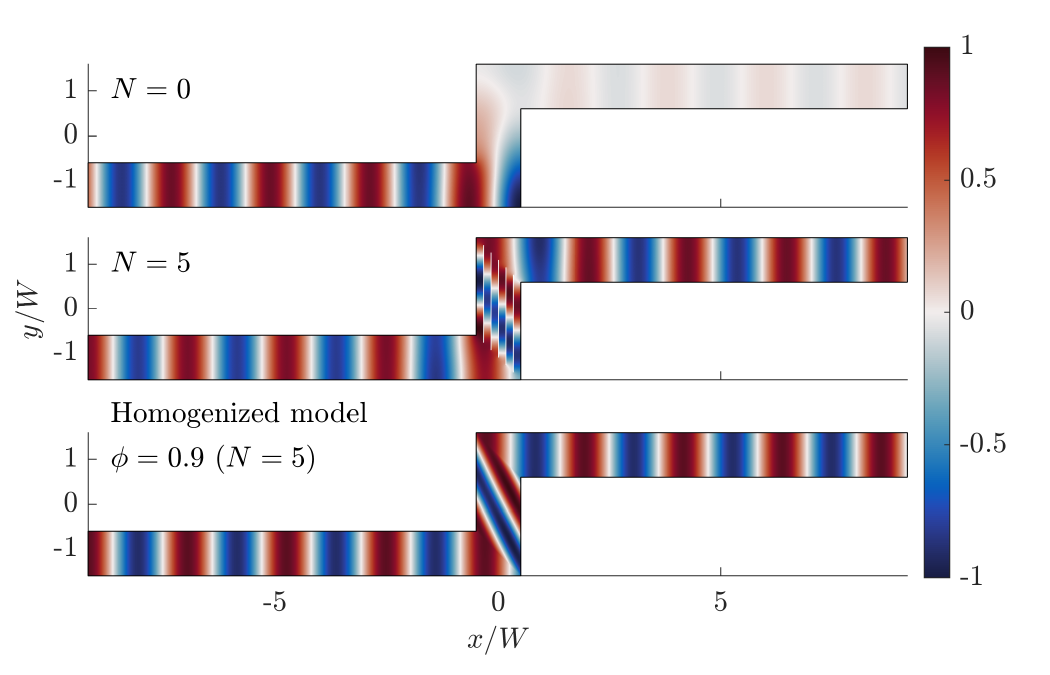}
    \caption{
    Numerical result: real part of the simulated field at $kW = 9\pi/10$ for an empty channel, a channel with five vertical plates in the middle region, and a homogenized medium with the filling ratio corresponding to $N = 5$.
    } 
    \label{fig:fields_simulation} 
\end{figure}

As stated previously, the plate-array metamaterial only allows 1D propagation in the direction of the plates, since the periodicity of the array is much smaller than the wavelength. It has been shown that, under this approximation, the periodic structure yields the same wave properties as an effective medium obtained through homogenization theory \cite{Norris2015, Marigo2017}. Then, the plate-array metamaterial can be replaced by a homogeneous anisotropic medium and the homogenized wave equation takes the form

\begin{equation}
	\text{div}\mathbf{U} + \phi k^2\eta(x,y) = 0, \quad 
	\mathbf{U} = \Big(
 	\begin{matrix}
 	0 & 0 \\
 	0 & \phi 
 	\end{matrix} \Big) \nabla \eta(x,y)
	\label{eq:homog}	
\end{equation}

\noindent with $\eta$ and $\mathbf{U \cdot n}$ being continuous at the interfaces between the channel and the effective medium, and where $\phi$ is the filling ratio of water, given by 

\begin{equation}
	\phi = 1-Nd/W.
	\label{eq:filling_ratio}
\end{equation}
\noindent $\phi=0$ corresponds to the entire channel filled with plates, while $\phi=1$ corresponds to plates with zero thickness.

In order to get more insight into the modelling of the behavior of the system, we are going to compare 2D numerical simulations of the channel with the plate-array metamaterial with the homogenization results. To illustrate this comparison, Fig. \ref{fig:fields_simulation} shows the free surface deformation field obtained numerically for a fixed frequency for the channel in three different configurations:  without the metamaterial, with a metamaterial made up of $N = 5$ vertical plates and the corresponding effective medium with $\phi = 0.9$. We observe that, as already mentioned, there is a very high reflection in the absence of the metamaterial, while the backscattering is considerably reduced when the plates or the effective medium are added. We also note a very good agreement between the homogenized model and the real geometry (as also shown in the Appendix B). This agreement improves as the system approaches the solution of an infinite array of infinitesimally thin plates, i.e. by increasing the number of plates and decreasing their thickness. 

\begin{figure}[!t] 
    \centering 
    \includegraphics[width=\linewidth]{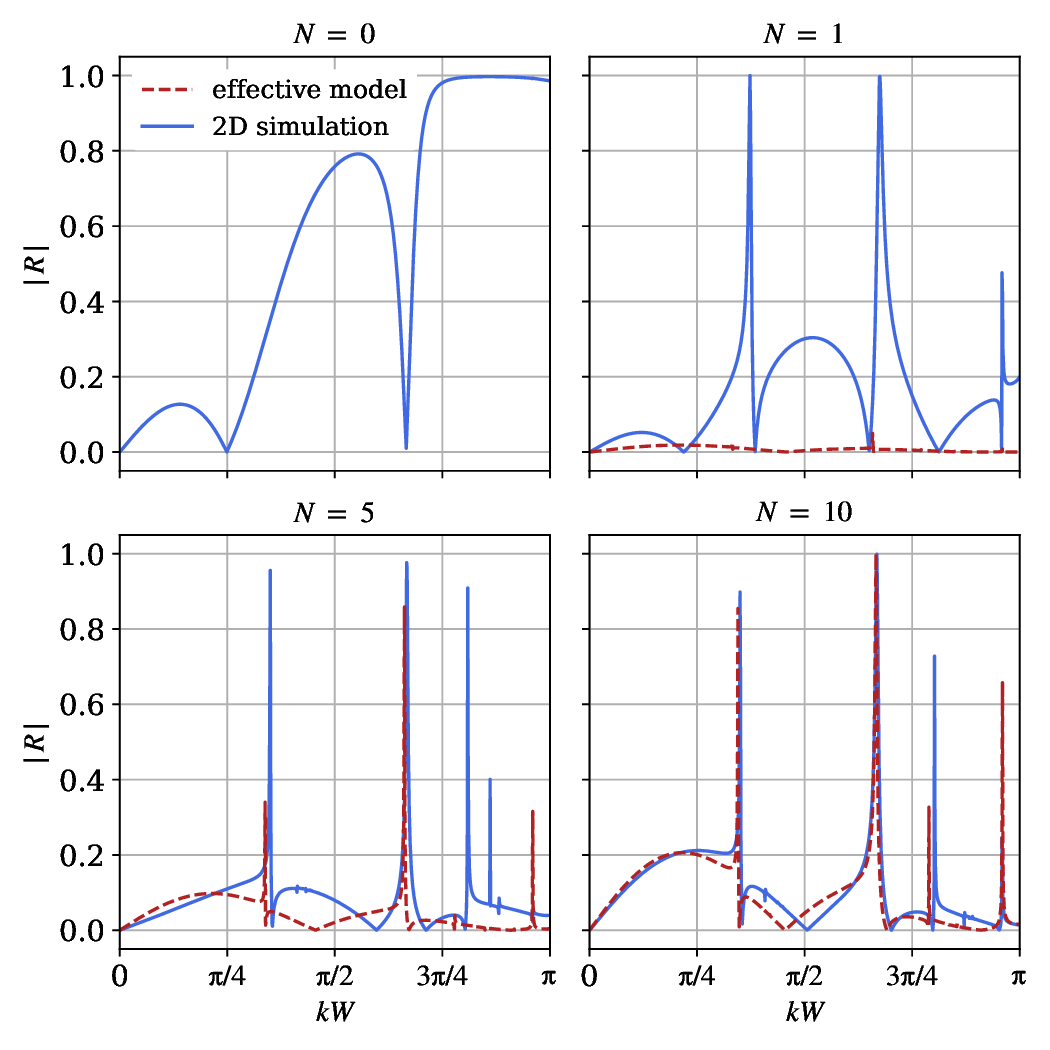}
    \caption{Reflection coefficient for different numbers of plates with a fixed thickness of $d = 0.02 W$ (blue solid line) and the one obtained from the effective model with the corresponding value of the filling ratio (red dashed line).
    } 
    \label{fig:R_vs_kw_multiple_N} 
\end{figure}

The filling ratio of water can be adjusted by either changing the thickness of the plates or their quantity. In this study, the plate thickness is fixed at $d/W=0.02$, as it strikes a balance between minimizing backscattering and enabling practical experiments. Therefore, we varied the number of plates $N$ to modify the filling ratio. The reflection coefficient was then computed at various frequencies for different values of $N$, as illustrated in Fig. \ref{fig:R_vs_kw_multiple_N}. The reflection coefficient obtained from the homogenized model is also depicted in the same figure, with the corresponding value of $\phi$ for each $N$, related by eq. \ref{eq:filling_ratio}.

In Fig. \ref{fig:R_vs_kw_multiple_N} we can observe that, in the absence of plates ($N=0$), $|R|$ increases with the frequency and exhibits the Fabry-Perot resonance, as seen in Fig. \ref{fig:fabry-perot}. Surprisingly and interestingly, the addition of only one plate already changes drastically the behavior of $|R|$, resulting in low reflection values across all frequencies, except at the resonances. In that case, of course, the effective medium does not describe the system accurately. The reflection coefficient obtained from the homogenized model converges to the one from the actual problem as the number of plates increases (i.e. when the filling ratio is closer to one). It is worth mentioning that the results of the 2D simulations with plates were used to average across all frequencies and obtain the data points presented in Fig. \ref{fig:optimal_n_multiple_thickness}.

\section{\label{sec:experiments}Experimental results}

\begin{figure}[!t] 
    \centering 
    \includegraphics[width=\linewidth,trim={0.7cm 4cm 0 0},clip]{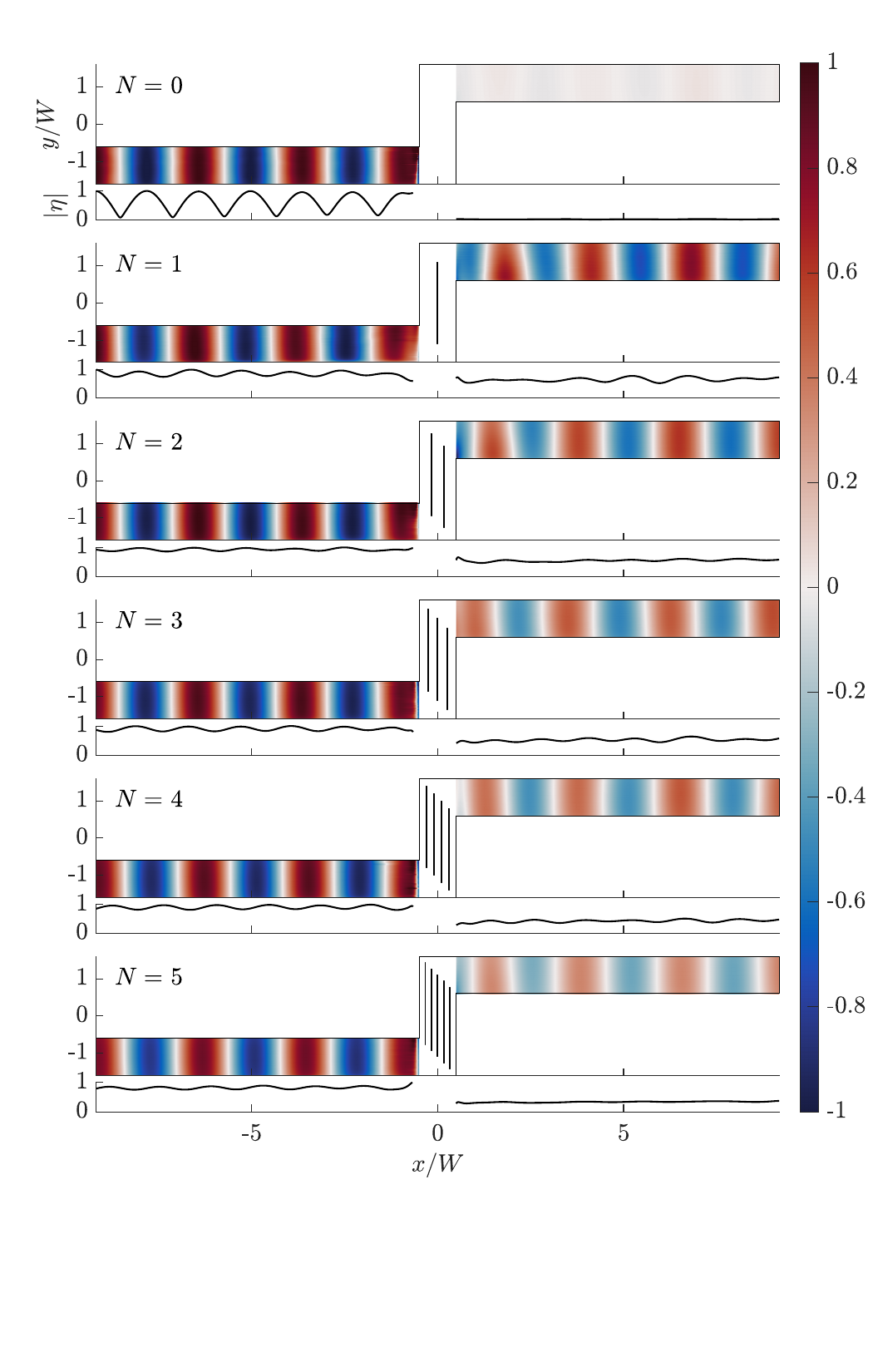}
    \caption{
    Experimental result: real part (top) and average of the absolute value (bottom) of the measured free surface deformation filtered at the forcing frequency for $kW = 9\pi/10$  (normalized).
    } 
    \label{fig:filtered_fields} 
\end{figure}

\begin{figure}[!t] 
    \centering 
    \includegraphics[width=\linewidth, trim={0.3cm 0 0.3cm 0},clip]{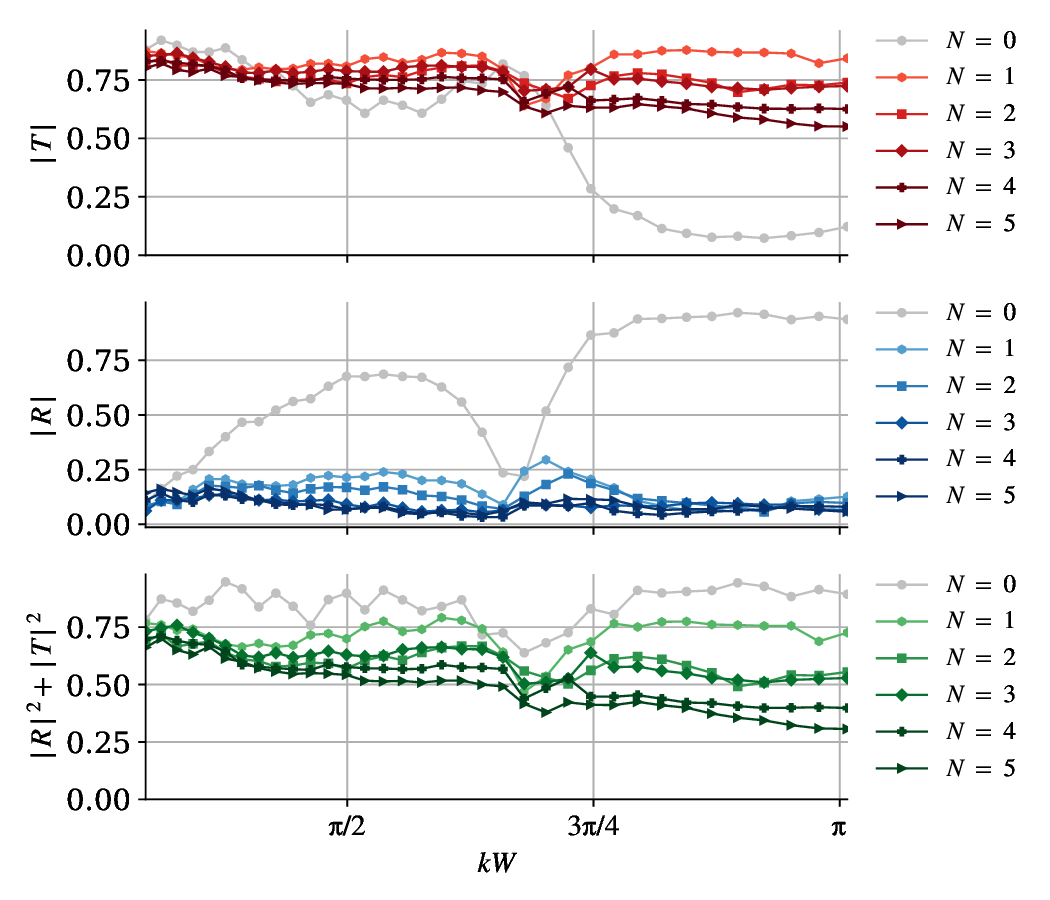}
    \caption{
     Reflection and transmission coefficients obtained experimentally for different numbers of plates.
    } 
    \label{fig:RTE} 
\end{figure}

Experiments were performed in a 2.95 m long channel of width $W$ = 0.10 m, with a \MODIF{turn} of $L_0$ = 0.32 m and a constant water level of $h$ = 0.05 m. Waves are generated at one side of the channel by a wavemaker for frequencies in the range $f \in$ [1.0, 2.7]~Hz, with a 0.05 Hz step ($f_{\text{cut-off}} = 2.69$ Hz). The frequency is linked to the wavenumber $k$  through the dispersion relation of gravity-capillarity waves

\begin{equation}
	\omega^2 = \Big( gk + \frac{\sigma}{\rho}k^3\Big)\tanh kh,
\end{equation}
where $\omega = 2 \pi f$, $g$ is the gravity acceleration, and where $\rho$~=~1000~kg/m$^3$  and $\sigma$ = 71 mN/m are respectively the water density and surface tension. \MODIF{This equation is only used to link the frequency $\omega$ imposed by the wavemaker with its corresponding number $k$.}


The wavemaker is driven by a linear motor that generates a sinusoidal vertical movement with an amplitude of 2 mm.  This amplitude has been checked to be low enough to remain in the linear regime. On the other side of the channel we placed an absorbing beach inclined at 5$^{\circ}$ in order to reduce spurious reflections from the outgoing region. Measurements were taken using the Fourier Transform Profilometry method \cite{Cobelli2009, Maurel2009c}, that allows single-shot measurements of the whole free surface. A sinusoidal (reference) pattern is projected on the free surface of the liquid and this projection is filmed from above. We used distilled water with a small amount of TiO$_2$ (4 g/l) in order to be able to project on its surface; it has been shown that this concentration does not modify significantly the hydrodynamical properties of water \cite{Przadka2012a}. 
The phase of the projected pattern is modified by the deformations of the free surface; the difference between this deformed phase and the reference one allows for the reconstruction of the height field. Two identical systems consisting  of a video projector (Epson EH-TW9400) and a USB camera (Basler acA1920-155um) arranged in a parallel-axis geometry were mounted and synchronized at each side of the perpendicular \MODIF{turn}. We took measurements of the full width of the channel, 1.12~m at each side of the \MODIF{turn}, at 30 fps.

We investigated experimentally the effect of the filling ratio, changing the number of plates with a fixed thickness of $d$ = 2 mm. Fig. \ref{fig:filtered_fields} shows the height fields at $kW = 9\pi/10$ for different numbers of plates between $N=0$ ($\phi=1$) and $N = 0$ ($\phi = 0.90$). In the absence of metameterial, the transmission is close to zero, while in its presence it increases considerably. As mentioned earlier with the numerical results, it is worth noting that the addition of only one plate has already a remarkable effect. 

To obtain a more comprehensive understanding than just at a single frequency, we subsequently explored the broadbandness of the backscattering reduction.
In order to extract the reflection and transmission coefficients, we filtered the height field at the forcing frequency and averaged it in the $y$-direction: $\eta(x) = \langle \eta(x,y) \rangle_{y}$. Then, since only the plane mode is propagating, $\eta$ can be written as
\begin{align}
	\eta(x) = 
	\begin{cases}
		A(e^{ikx} + Re^{-ikx}), \quad & \text{in region I},\\
		ATe^{ikx}             , \quad & \text{in region II},
	\end{cases}
	\label{eq:fit}
\end{align}
\noindent where $A$ is the amplitude of the incident wave, $R$ and $T$ the reflection and transmission coefficients, and regions I and II correspond respectively to the far field before and after the \MODIF{turn} (i.e., for $x<0$ and $x>0$). Fig. \ref{fig:RTE} shows the reflection and transmission coefficients as well as the energy for 35 different frequencies, obtained by performing a fit on each side with eq. \ref{eq:fit} \cite{Bobinski2018}. For the measurement with $N = 0$ (gray line), we recover the behavior of the reflection and transmission coefficients predicted numerically as well as the Fabry-Perot resonance, which disappears almost completely by the addition of the plates. Besides, we note that the scattered energy flux decreases as the number of plates increases for all frequencies. \MODIF{This is due to losses in the experiment that were not considered in the numerical approach. If we consider only viscous losses given by the bulk viscosity and the bottom and wall friction, the theory \cite{gutierrez2017experimental,hunt1952viscous,mei1973damping} predicts much lower dissipation than the one observed in the experiment (Fig. \ref{fig:RTE}). Therefore, we can reasonably assess that this dissipation is generated mostly by meniscus effects.}
Consequently, the optimal number of plates does not match the one found numerically, since the transmission past the \MODIF{turn} diminishes with the number of plates. However, even in the cases where high transmission is not well achieved, the plate-array reduces the backscattering significantly. 

With a detailed inspection of Fig. \ref{fig:RTE}, we can distinguish several behaviors: the plate-array metamaterial can yield high transmission, low reflection or both, depending on the parameters chosen (filling ratio and frequency). When adding the metamaterial, the reflection drops drastically and takes values lower than 30\% for all tested configurations, being progressively smaller as the number of plates increases. Nevertheless, due to the viscous losses, an experimental minimum of reflection does not necessarily correspond to a maximum of transmission. Indeed, the maximum of transmission occurs when $N = 1$; this case is of particular interest especially for high frequencies, where, as it turns out, adding as few as only one plate is enough to considerably increase the transmission through the turns.

\section{\label{sec:conclusions}Concluding remarks}
We designed and built an experimental plate-array metamaterial capable of reducing the backscattering of surface waves in a \MODIF{sharply bent} waveguide for a broadband range of frequencies without relying on resonant phenomena. Essentially, the role of the metamaterial is to reduce the reflection by impedance matching and, in a lossless case, it implies automatically a high transmission. 
In an experimental realization, a high attenuation due to viscous friction might be anticipated since the plates are closely spaced. However, by finely tuning the number of plates, we have shown that this device not only reduces the backscattering but also allows for a remarkable transmission. Surprisingly, even using only one plate provides already a very strong reduction of backscattering and a good transmission.  
It is important to point out that the broadband backscattering reduction can be achieved for turns in channels with angles other than $\pi/2$, where it can lead to the construction of more complex devices, e.g. with several successive turns.

\begin{acknowledgments}
S.K., P.P., A.M. and V.P.  acknowledge the support of the Agence Nationale de la Recherche (ANR) under grant 243560 CoProMM.
\end{acknowledgments}

\appendix
\section*{\label{sec:fabry_perot}Appendix A: 1D model} 
\begin{figure}[!t] 
    \centering 
    \includegraphics[width=0.9\linewidth,  trim={0 3cm 0 2cm},clip]{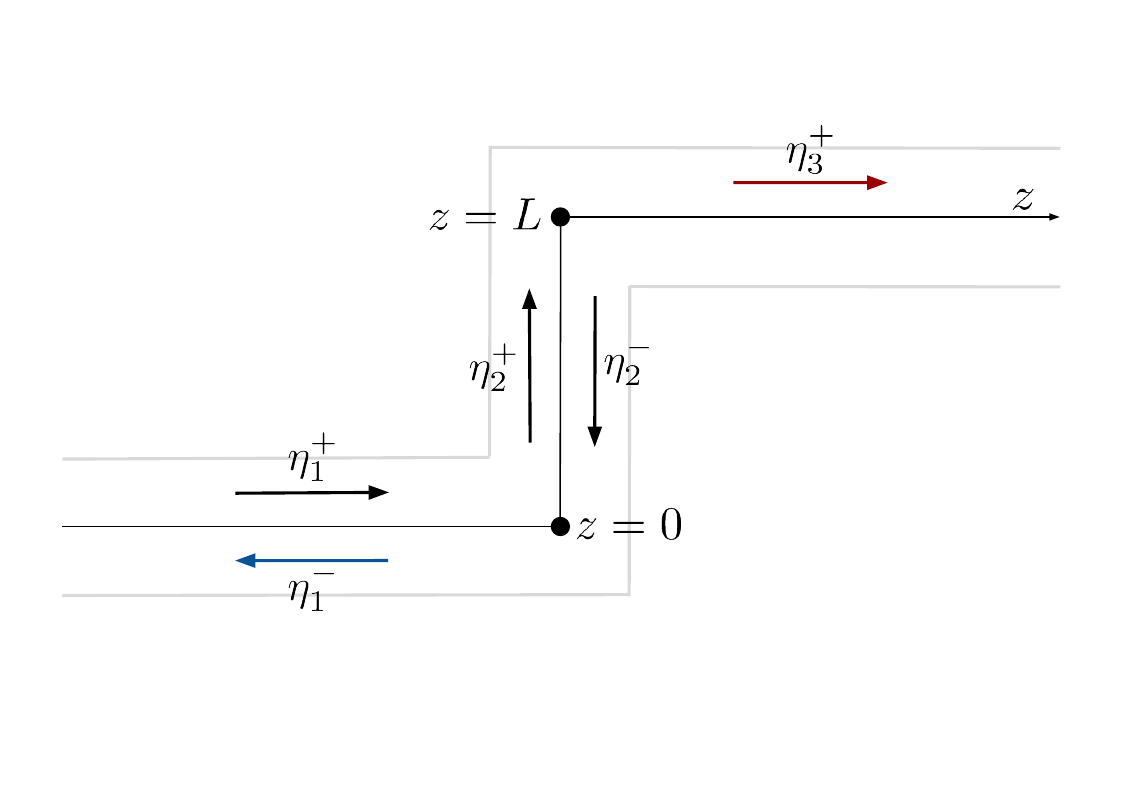}
    \caption{Propagative waves established in the stationary solution for a 1D problem with two reflective interfaces.} 
    \label{fig:appendix} 
\end{figure}

We consider the problem of a plane wave within time harmonic regime that passes through two refractive interfaces separated by a distance $L$. When the incident wave reaches the first interface, there is a reflected and a transmitted wave. The latter is then reflected and transmitted at the second interface. The reflected wave coming from the second interface arrives again at the first one, and this process continues an infinite number of times. Therefore, is necessary to consider the infinite sum of reflected and transmitted waves. Eventually, after the infinite reflections, a stationary solution consisting of five propagating wave components is established, as shown in Fig. \ref{fig:appendix}. 

These five waves can be expressed as
\begin{equation} \label{eq:waves_def}
\begin{split}
	\eta_1^+(z) & = A_1e^{ikz} \\
	\eta_1^-(z) & = B_1e^{-ikz} \\
	\eta_2^+(z) & = A_2e^{ikz} \\
	\eta_2^-(z) & = B_2e^{-ikz} \\
	\eta_3^+(z) & = A_3e^{ikz},
\end{split}
\end{equation}

\noindent where $A_1$ is known, $A_1, A_2, A_3, B_1, B_2 \in \mathbb{C}$ and they can be obtained from the matching conditions at $z=0$ and $z~=~L$. We will express them as a function of the reflection and transmission coefficients of the interfaces, $r$ and $t$, supposed to be symmetric and the same for both of them.
\begin{equation} \label{eq:bc}
\begin{split}
	\eta_2^+(z=0) & = t\eta_1^+(z=0) + r \eta_2^+(z=0) \\
	\eta_1^-(z=0) & = r\eta_1^+(z=0) + t \eta_2^-(z=0) \\
	\eta_2^-(z=L) & = r\eta_2^+(z=L) \\
	\eta_3^+(z=L) & = t\eta_2^+(z=L)
\end{split}
\end{equation}

Evaluating the expressions at $z=0$ and $z=L$, we obtain
\begin{equation} \label{eq:bc_amp}
\begin{split}
	A_2 & = tA_1 + rB_1 \\
	B_1 & = rA_1 + t B_2 \\
	B_2 e^{-ikL} & = rA_2 e^{ikL} \\
	A_3 e^{ikL} & = tA_2 e^{ikL}.
\end{split}
\end{equation}

The amplitude of the outgoing wave $\eta_3^+$ can be expressed as
\begin{equation}
	A_3 = \frac{t^2 A_1}{1-r^2e^{2ikL}}.
\end{equation}

This expression allows to calculate the transmission as
\begin{equation}
	| T | ^2 = \frac{| A_3 | ^2 }{| A_1 | ^2}= \frac{|t|^4}{|1-r^2e^{2ikL}|^2}.
\end{equation}

\noindent Noting $r = r_0e^{i\varphi}$ (with $r_0 = |r|$) and using the property $|r|^2 + |t|^2 = 1$,  we can write $|T|^2$ as
\begin{equation}
	|T|^2 = \frac{1}{1+F\sin^2(\frac{\delta}{2})},
\end{equation}

\noindent where  $F = 4r_0^2/(1-r_0^2)^2$ and $\delta = 2kL+2\varphi$. This equation depends on the absolute value of the reflection coefficient $r_0$ and its phase $\varphi$, as well as the distance $L$ between the interfaces. Since $|R|^2 + |T|^2 = 1$, we can easily obtain the reflection coefficient:
\begin{equation}
	|R|^2 = \frac{F\sin^2(\frac{\delta}{2})}{1+F\sin^2(\frac{\delta}{2})}.
\end{equation}

\section*{\label{sec:one_corner}Appendix B: One corner}

\begin{figure}[!t] 
    \centering 
    \includegraphics[width=\linewidth,  trim={0 0 0 1cm},clip]{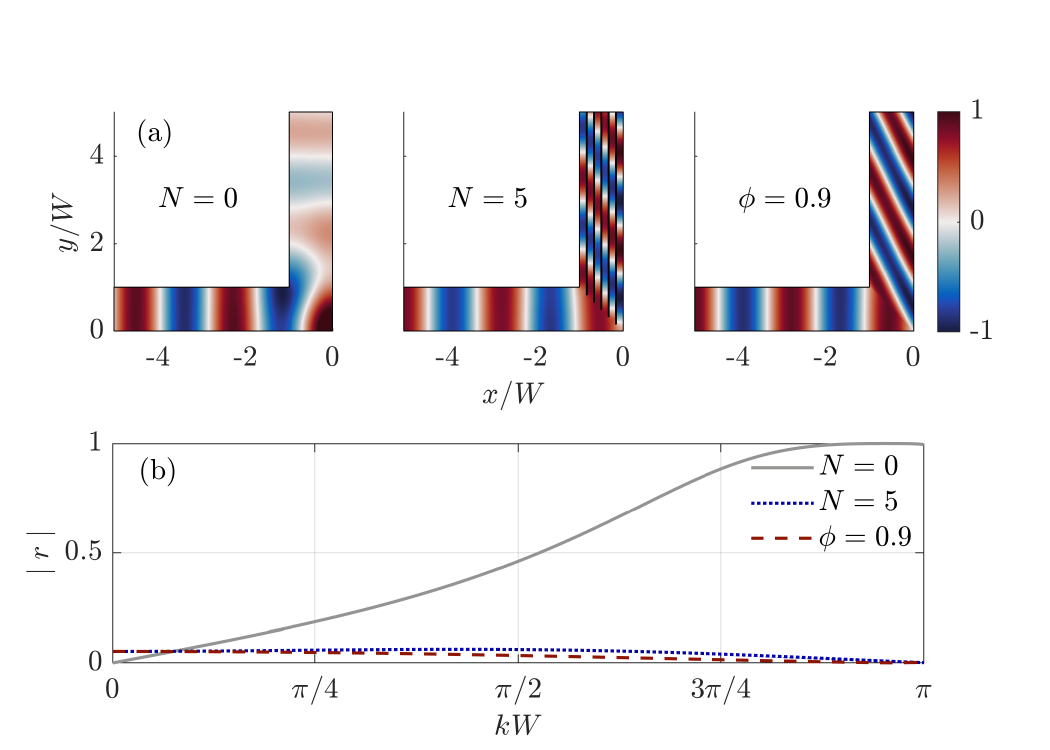}
    \caption{Numerical result: (a) Real part of the simulated field at $kW = 9 \pi /10 $ for an incident wave reaching one perpendicular turn with no plates, five vertical plates and a homogenized medium with the filling ratio corresponding to $N = 5$. (b) Absolute value of the reflection coefficient for the above configurations.}
    \label{fig:appendix_corner} 
\end{figure}

We consider an incident plane wave with harmonic time dependence propagating in a waveguide of width $W$ and reaching one perpendicular turn. We consider frequencies below the cut-off frequency, i.e., $kW < \pi$, where only the planar mode can propagate. We performed numerical simulations in order to find the reflection coefficient $r$ for this system.  Three different configurations were considered: an empty channel, one with a metamaterial made of $N = 5$ plates after the turn, and one with a homogenized medium in its place, using the filling ratio $\phi=0.9$ that corresponds to $N=5$ plates. Fig. \ref{fig:appendix_corner} shows the simulated fields as well as the absolute value of $r$ as a function of the frequency for each case. While $|r|$ increases with frequency for the case without plates, it remains mostly constant and near zero for the other two cases (with $N=5$ plates and a homogenized medium with $\phi = 0.9$), as expected. It should be also noted that there is good agreement between the homogenized model and the field obtained with the plate-array metamaterial. The complex value of $r = r_0e^{i\varphi}$ for $N = 0$ was used to calculate the reflection coefficient $R$ of the complete system in the 1D model (eq. \ref{eq:fabryperot}).

\section*{\label{sec:infinitesimally_thin}\protect\MODIF{Appendix C: Infinitesimally thin plates}}

\begin{figure*}
    \centering 
    \includegraphics[width=\linewidth, trim={2cm 7cm 2cm 2cm},clip]{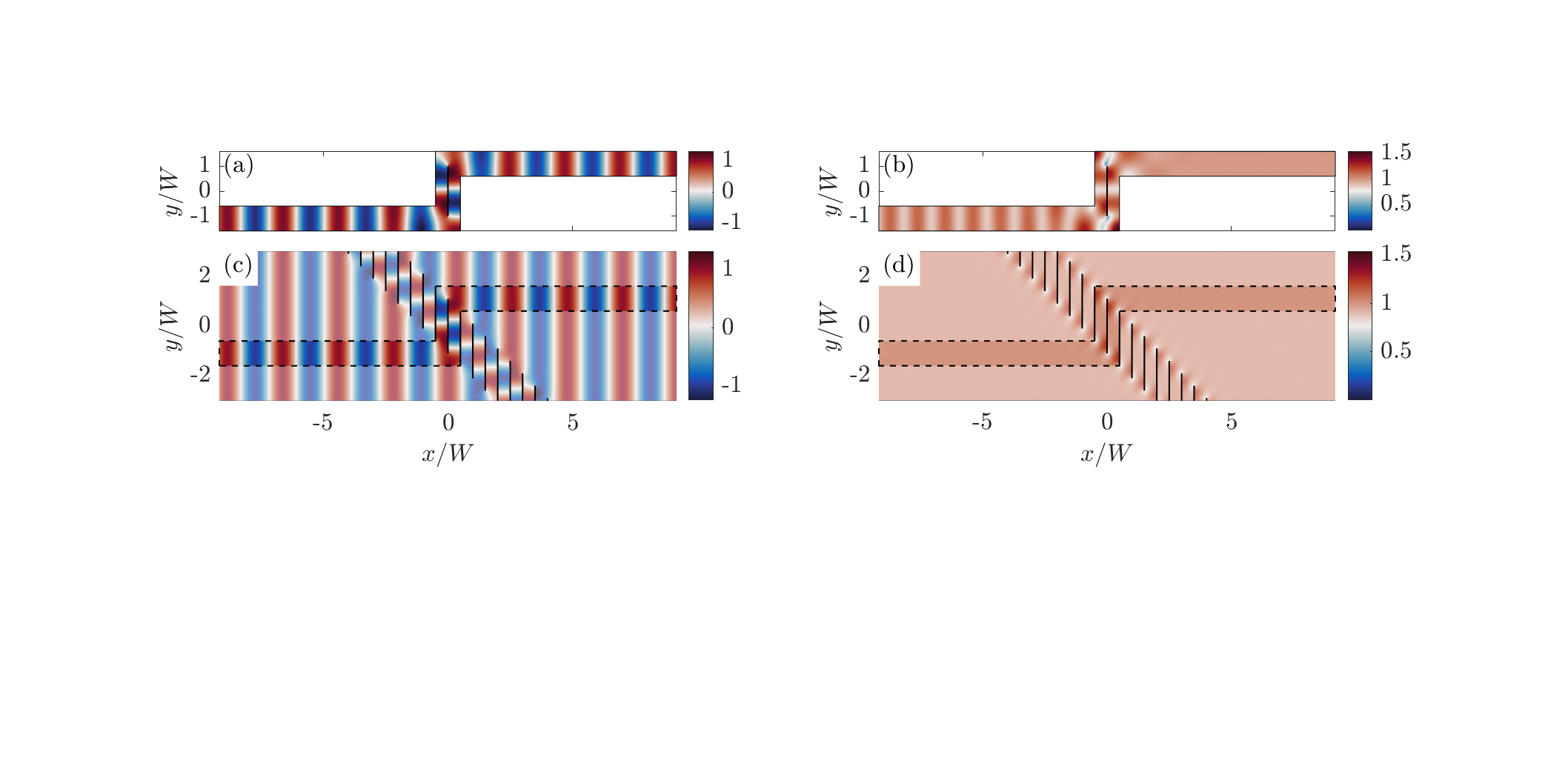}
    \caption{
    (a,c) real part and (b,d) absolute value of the free surface deformation at $kW = 9\pi/10$. (a,b) show the field in a waveguide with one infinitesimally thin plate, while (c,d) show the plane wave incidence on a periodic grating made of vertical plates of zero thickness. The grating has a periodicity of $a=W/2$. The dashed lines indicate where the channel walls would be if we confined this system into a waveguide, and the outer field is slightly masked.
    } 
    \label{fig:appendix_thin_plates} 
\end{figure*}

\MODIF{
	We consider an incident plane wave propagating in a bent channel with one infinitesimally thin plate in the central region and we compare it to a plane wave incidence on a periodic grating made of infinitesimally thin plates. The grating has a periodicity of $a = W/2$ to maintain the same distance between plates as in the case of a waveguide.}	

\begin{figure}[t] 
    \centering 
    \includegraphics[width=\linewidth, trim={1cm 0cm 1cm 0cm},clip]{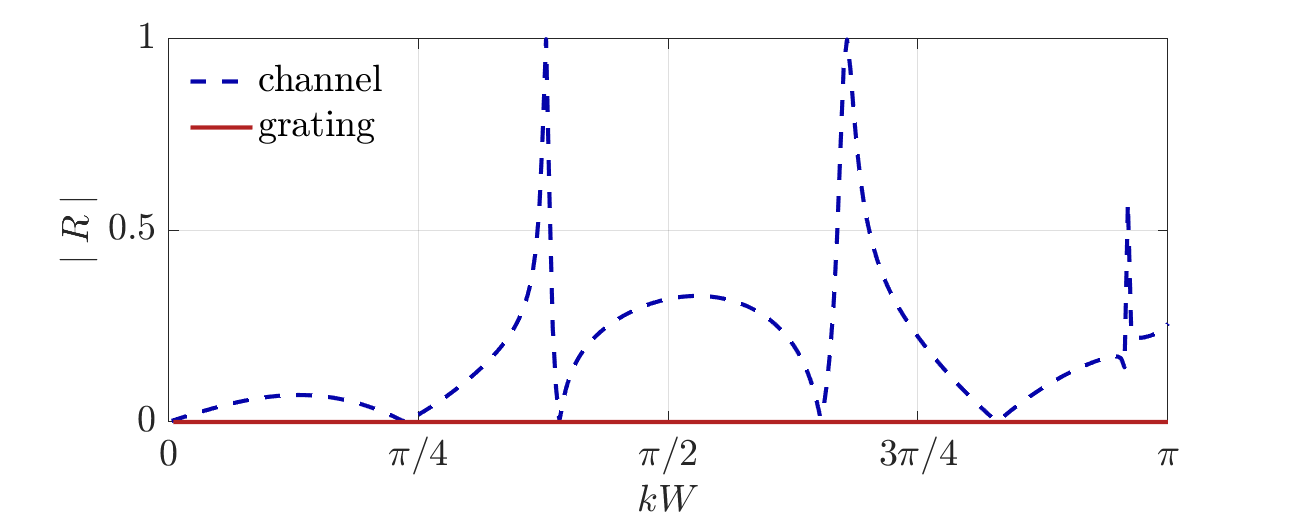}
    \caption{
	    Numerical result: absolute value of the reflection coefficient for a channel with one vertical plate of zero thickness (Fig \ref{fig:appendix_thin_plates}(a,b)) and for a periodic array of plates with zero thickness (Fig. \ref{fig:appendix_thin_plates}(c,d)).
    } 
    \label{fig:R_channel_thin} 
\end{figure}

\MODIF{
	Fig. \ref{fig:appendix_thin_plates} shows the real part and the absolute value of the free surface deformation field for a fixed frequency, both in a waveguide and in a periodic medium. Although these two cases might appear similar, there is a difference if we inspect closely in the neighborhood of the plates. The dashed lines in Fig. \ref{fig:appendix_thin_plates}(c,d) indicate where the channel walls would be if we confined this system into a waveguide. The presence of the channel imposes Neumann boundary conditions on the walls. This confinement leads to a different behavior of the field. In particular, the reflection coefficient for a plane wave incidence in a waveguide is no longer zero as in the case of a periodic grating, as shown in Fig. \ref{fig:R_channel_thin}.  
}

\bibliography{references.bib}

\begin{thebibliography}{36}
\bibitem [{not()}]{note_axis_grating}%
  \BibitemOpen
  \href@noop {} {}\bibinfo {note} {Note that the coordinate system
  {($\tilde{x}, \tilde{y}$)} is rotated by an angle {$\alpha$} with respect to
  {($x,y$)} from Fig. \ref{fig:setup}, so that the {$\tilde{y}$}-axis aligns
  with the interface between the region of the incident wave and the region
  occupied by the plates}\BibitemShut {NoStop}%
\end{thebibliography}

\end{document}